\documentclass[a4paper,twocolums]{article}
\usepackage{graphicx}
\usepackage{subfigure}
\graphicspath{{figs/}}
\usepackage{xcolor}
\usepackage{soul}
\begin{document}
\title{Energy and water vapor transport across a simplified cloud-clear air interface}

\author{L.\ Gallana, S.\ Di Savino, F.\ De Santi, M.\ Iovieno, D.\ Tordella}
\maketitle
\begin{abstract}
We consider a simplified physics of the could interface where condensation, evaporation and radiation are neglected and  momentum, thermal energy and water vapor transport is represented in terms of the Boussinesq model coupled to a passive scalar transport equation for the vapor. The interface is modeled as a layer separating two isotropic turbulent regions with different kinetic energy and vapor concentration. In particular, we focus on the small scale part of the inertial range as well as on the dissipative range of scales which are important to the micro-physics of warm clouds. We have numerically investigated stably stratified interfaces by locally perturbing at an initial instant the standard temperature lapse rate at the cloud interface and then observing the temporal evolution of the system. 
When the buoyancy term becomes of the same order of the inertial one, we observe a spatial redistribution of the kinetic energy which produce a concomitant pit of kinetic energy within the mixing layer.
In this situation, the mixing layer contains two interfacial regions with opposite kinetic energy gradient, which in turn produces two  intermittent sublayers in the velocity fluctuations field. This changes the structure of the field with respect to the corresponding non-stratified shearless mixing: the communication between the two turbulent region is weak, and the growth of the mixing layer stops. These results are discussed {with respect to experimental} results with and without stratification.
\end{abstract}

\section{Introduction}
Warm clouds as stratocumuli swathe a significant part of earth's surface and play a major role in the global dynamics of atmosphere by strongly reflecting incoming solar radiation -- thus contributing to the Earth's albedo -- so that an accurate representation of their dynamics is important in large-scale analyses of atmoshperic flows \cite{wo12}.
They are controlled by the tight interplay between radiative driving, turbulence, surface fluxes, latent heat release, and entrainment. Among them, the mixing and entrainment processes at the cloud top have been identified as fundamental to determine the internal structure of warm clouds, so that a clear and complete understanding of their physics is required \cite{ger2013}. 
As pointed out by Malinowsky \textsl{et al.} \cite{mal2013}, data from most field campaigns and large-eddy simulations are too poorly resolved to allow to infer the details of the interfacial layer, even if they indicate that, in order to allow for entrainment, a high level of turbulence must be present.
For this reason, in this work we study the local transport through a clear air/cloud interface through DNS (Direct Numerical Simulation). As our focus is on the dynamics of the smallest scales of the flow which influence the microphysics of warm clouds, we have simulated an idealized configuration to understand, under controlled conditions, some of the basic phenomena which occur at the cloud interface over length scales of the order of few meters. In these conditions, we solve scales from few meters down to few millimeters, that is, we resolve only the small scale part of the inertial range and the dissipative range of the power spectrum in a small portion ($6\ m\times6\ m\times12\ m$) of the atmosphere across the clou - clear air interface.
This allows us to investigate the dynamics of entrainment which occurs in a thin layer at the cloud top, which a smaller scale with respect ot the scale explicitly resolved in large-eddy simulations of clouds \cite{moe2000}. 
In this preliminary work, we focus on two concomitant aspects of the cloud top mixing layer: the effect of the presence of a stratification and of a turbulent kinetic energy gradient. We do not consider the wind shear neither the phenomena linked to the processes of evaporation and condensation and radiative cooling which are important in conditions of buoyancy reversal \cite{mel2010, mel2014}.
Therefore, our simulations were performed by applying the Boussinesq approximation to the Navier-Stokes momentum and energy equations together with a passive scalar transport equation which models the water vapor transport. 


The details of the physical problem we have considered and of the governing equations are given in section 2. Section 3 contains a selection of our main results about intermittency, energy redistribution and entrainment.
Conclusion remarks are in Section 4.

\section{The physical problem}
\begin{figure}
\centering
\begin{minipage}{.5\textwidth}
\includegraphics[width=\textwidth]{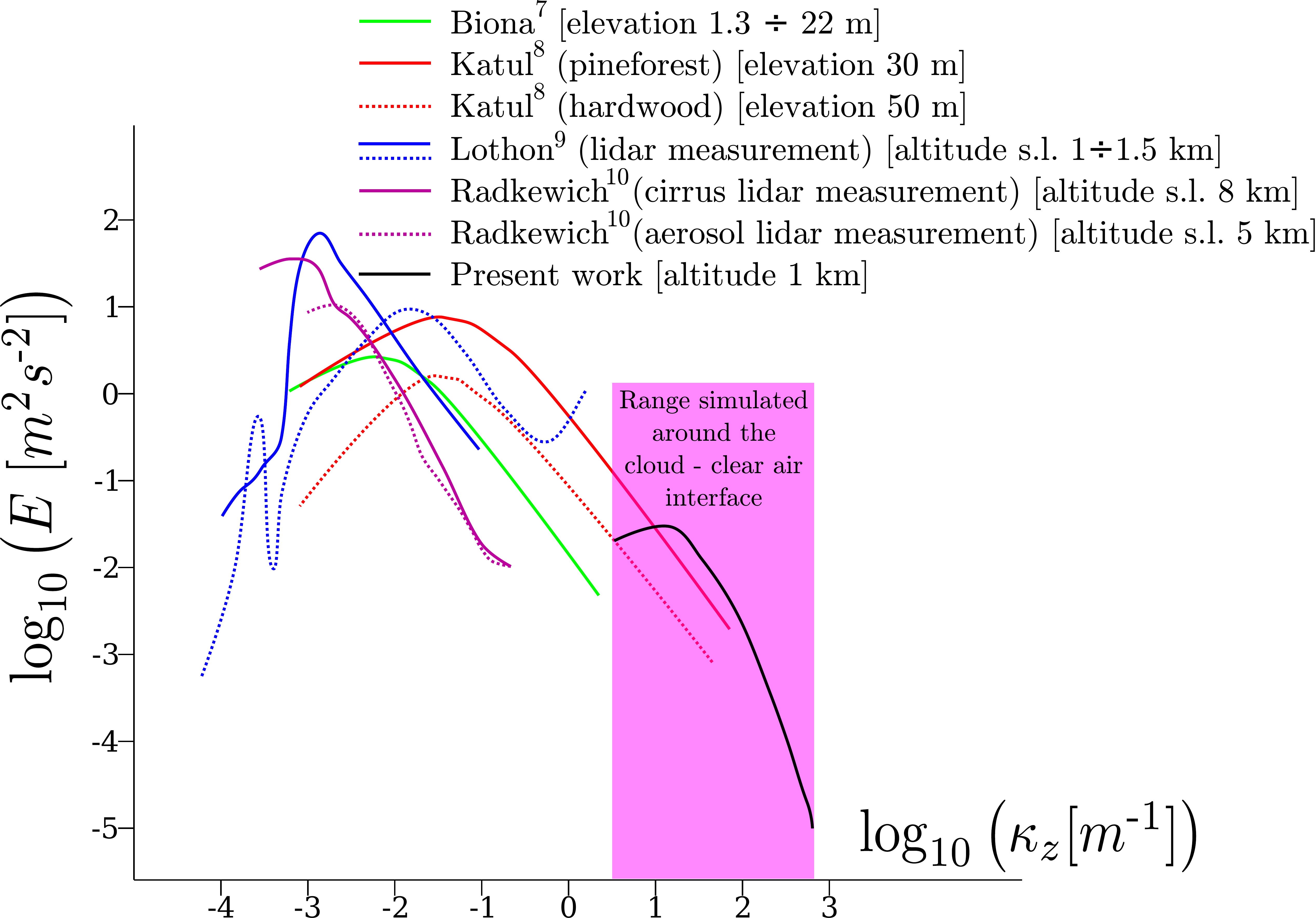}
\end{minipage}
\hfill
\begin{minipage}{.45\textwidth}
\caption{\label{fig:spett}{Kinetic energy spectra. Contextualization of present study (black spectra, small inertial and dissipative range) respect to spectra from in-situ atmospheric measurements \mbox{\cite{biona, katul, lot, radk}} (colored spectra,  energy injection and low wave-number inertial scales). }}
\end{minipage}
\end{figure}
\begin{figure}
\centering
\includegraphics[width=\textwidth]{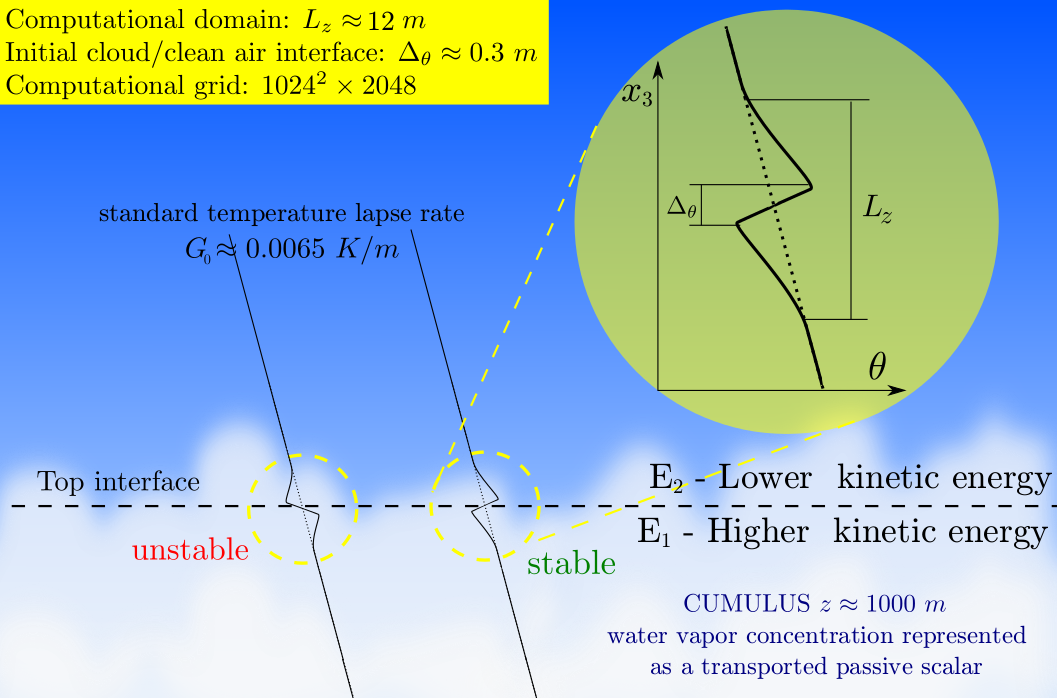}
\label{fig:schema}
\caption{Scheme of the initial conditions. $E_1$ is the mean initial turbulent kinetic energy of the bottom region (inside the cloud), $E_2$ of the upper region (outside the cloud). {For the top cloud mixing here presented we consider $E_1/E_2=6.7$}. The water vapor is initially present only inside the cloud (bottom region). The zoom in the yellow circle is an example of {initial temperature perturbation $\theta'$} of the standard boundary layer lapse rate.}
\end{figure}

We consider the interface between clear air and cloud in a $6\ m \times 6\ m\times12\ m$ parallelepipedic domain. {Compared to in situ measurement of the atmospheric energy spectra, as shown in figure \mbox{\ref{fig:spett}}}, we are able to simulate the lowest part of the inertial range and the dissipative one. As shown in figure \ref{fig:schema}, the system is composed by two homogeneous and isotropic turbulent regions that interact through a mixing layer, whose initial thickness has been set of the same order of {the integral scale of the turbulence background $\ell$}, here assumed equal to $3\cdot10^{-2}$ m. The two isotropic regions have a different kinetic energy and we assumed that the kinetic energy is higher in the cloud than in the external region. The root mean square of the velocity inside the cloud is $u_{rms}=0.2$ m/s, and {the energy ratio between the cloud energy {$E_1$} and the external region energy $E_2$ is equal to 6.7}. This energy ratio is of the same order of the ones measured in warm clouds (see, e.g., \cite{mal2013}) and, furthermore, it allows us to compare our results with experiments on shearless mixing (see \cite{vw89,prl11}) in absence of any stratification. In our simulations the Taylor microscale Reynolds number $Re_\lambda$ of the higher energy region is equal to 250. Buoyancy is taken into account through a local perturbation {$\theta'$ in the profile of temperature distribution $\theta$} inside the troposphere, and  {the water vapor concentration $\chi$} is considered as a passive scalar. The Prandtl (Pr$=0.74$) and the Schmidt number ($Sc=0.61$) considered refers to an altitude of $1000\ m$ s.l.

{The initial conditions for the temperature perturbation is described in \mbox{\ref{fig:schema}} and in table \mbox{\ref{tabone}}.} The ratio between inertial and buoyancy forces is expressed by the Froude number $Fr$, based on the maximum gradient within the initial interface, which ranges from 31.2 (negligible stratification) to 0.62 (strong stratification).

\begin{table}
\caption{\label{tabone}Initial stratification level parameters. $G$ is the maximum gradient of $\theta$, expressed in terms of the standard troposphere {lapse rate} $G_0=0.0065\ Km^{-1}$; $N_{ci}=\sqrt{\alpha \theta_0 g \frac{d\theta}{dx_3}}$ is the 
 characteristic Brunt-V\"ais\"al\"a\ frequency of initial condition. The Froude number Fr$=\frac{u'_{{rms}}}{N_{ic}\ell}$ 
 and the Reynolds Buoyancy number Re$_b=\frac{\varepsilon N^2}{\nu}$ give a measure of the order of magnitude of the buoyancy forces compared with the inertial terms {($\varepsilon$ is the initial energy dissipation rate, $\nu$ the kinematic viscosity)}.}

\begin{center}
\begin{tabular}{*{5}{l}}
ine                              
$G$ & $\Delta T\ [K]$ & $N_{ic}\ [s^{-1}]$  & Fr & {Re}$_b$\cr 
ine
2$G_0$ & 4.0e-3 & 2.13e-2 & 31.2 & 7 \\
30$G_0$& 6.0e-2 & 5.24e-2 & 12.7 & 112 \\
100$G_0$ & 2.0e-1 & 1.50e-1 & 4.4 & 273 \\
500$G_0$ & 1.0e0 & 3.35e-1 & 1.8 & 833\\
5000$G_0$ & 1.0e1 & 1.06e0 & 0.62 & 2635\\
ine
\end{tabular}
\end{center}
\end{table}

We use the continuity, momentum and energy balance equations within the Boussinesq approximation, which holds for small temperature variations \cite{drazin}:

\begin{equation}
\nabla\cdot\mathbf u' = 0
\label{eq:mass}
\end{equation}
\begin{equation}
 \frac{\partial \mathbf{u}'}{\partial t} + \left(\mathbf{u}'\cdot\nabla\right)\mathbf{u}' = -\nabla \frac{\tilde{p}}{\rho}+\nu \nabla^2\mathbf{u}' + \alpha\mathbf{g}\theta'
 \label{eq:mom2}
\end{equation}
\begin{equation}
 \frac{\partial \theta'}{\partial t} + \mathbf{u}'\cdot\nabla \theta'+u_3G = \kappa \nabla^2\theta'
 \label{eq:ene2}
\end{equation}
\begin{equation}
 \frac{\partial \chi}{\partial t} + \mathbf{u}'\cdot\nabla \chi = d_\chi \nabla^2\chi,
 \label{eq:scal}
\end{equation}
where the temperature $\theta$ is composed as the sum of a fluctuation $\theta'(\mathbf{x},t)$, a static component $\tilde{\theta}(x_3)=Gx_3$, and a reference constant temperature $\theta_0$, $\tilde{p}=p+\alpha g x_3\left(\theta_0+Gx_3/2\right)$ {is the total hydrodynamic pressure (where $p$ is the fluid-dynamic pressure, $\alpha$ the thermal expansion coefficient, $g$ the gravity acceleration)}, $u'$ is the velocity fluctuation and $\chi$ is the vapor concentration of the air - water vapor mixture present in the cloud, here considered as a passive scalar. {The constant $\kappa$ and $d_\chi$ are respectively the thermal and water vapor diffusivity.}

{The simulations are performed using our home produced computational code that implements a pseudospectral Fourier-Galerkin spatial discretization and an explicit low storage fourth order Runge-Kutta time integration scheme. Evaluation of non-linear (advective) terms is performed through the 3/2 de-aliased method \mbox{\cite{ict01}}.
The initial conditions for the velocity field are obtained by a linear matching of two different isotropic homogeneous turbulent fields (that are randomly generated, respecting physical conditions imposing spectra, solenoidality, integral scale and kinetic energy)\mbox{\cite{ti06}}. 
The grid has $1024\times1024\times2048$ points, and allows to capture all the turbulent scales from the greatest (integral scale $\ell$) to the smaller (Kolmogorov scale $\eta$). The computational code uses a distributed memory paradigm through the MPI libraries: the simulation were performed at the TGCC Curie supercomputer within the PRACE project n$^\circ$ RA07732011 for a total of 3 million cpu-hours.}

\section{Results}
\begin{figure}[tb]
	\centering
	\subfigure[]{\includegraphics[width=.49\textwidth]{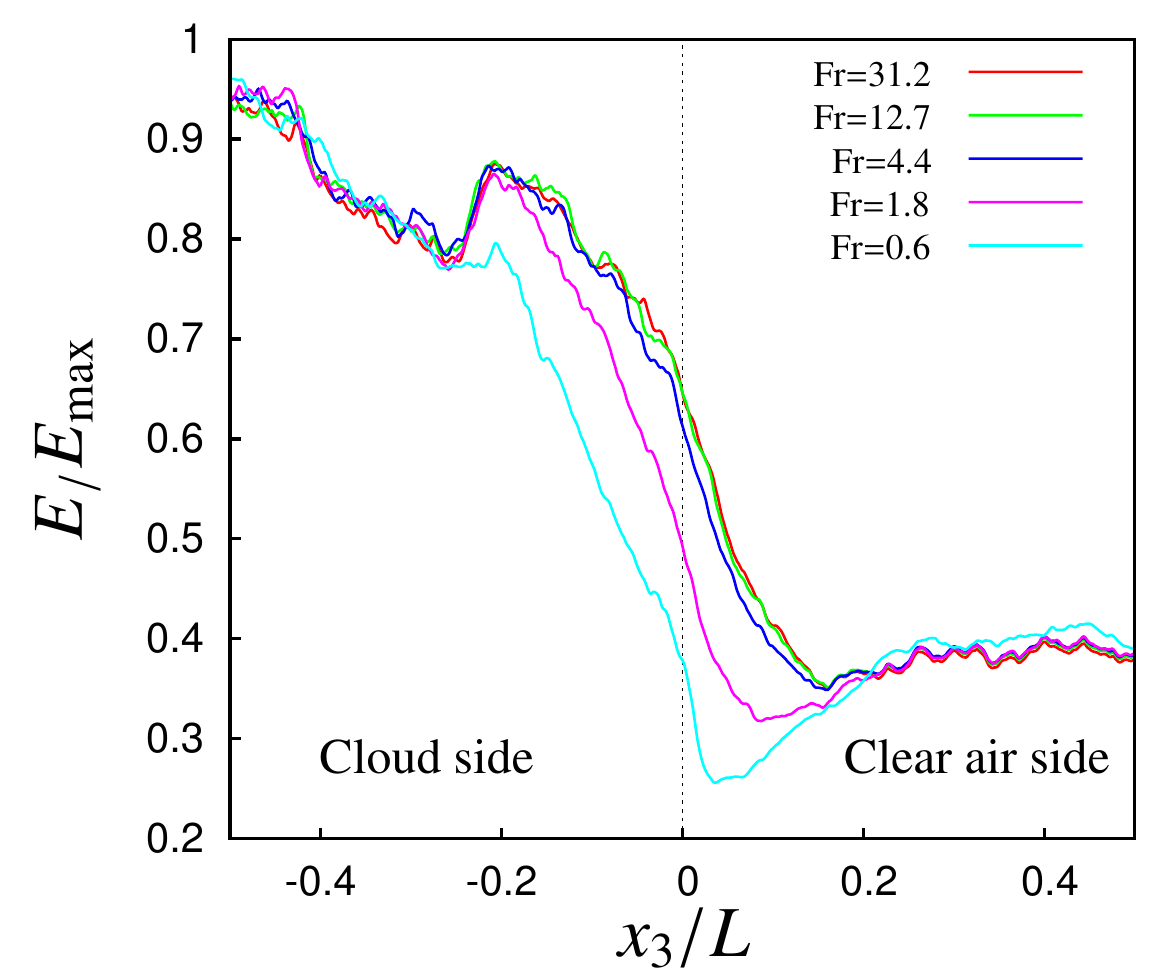}}
	\hfill
	\subfigure[]{\includegraphics[width=.49\textwidth]{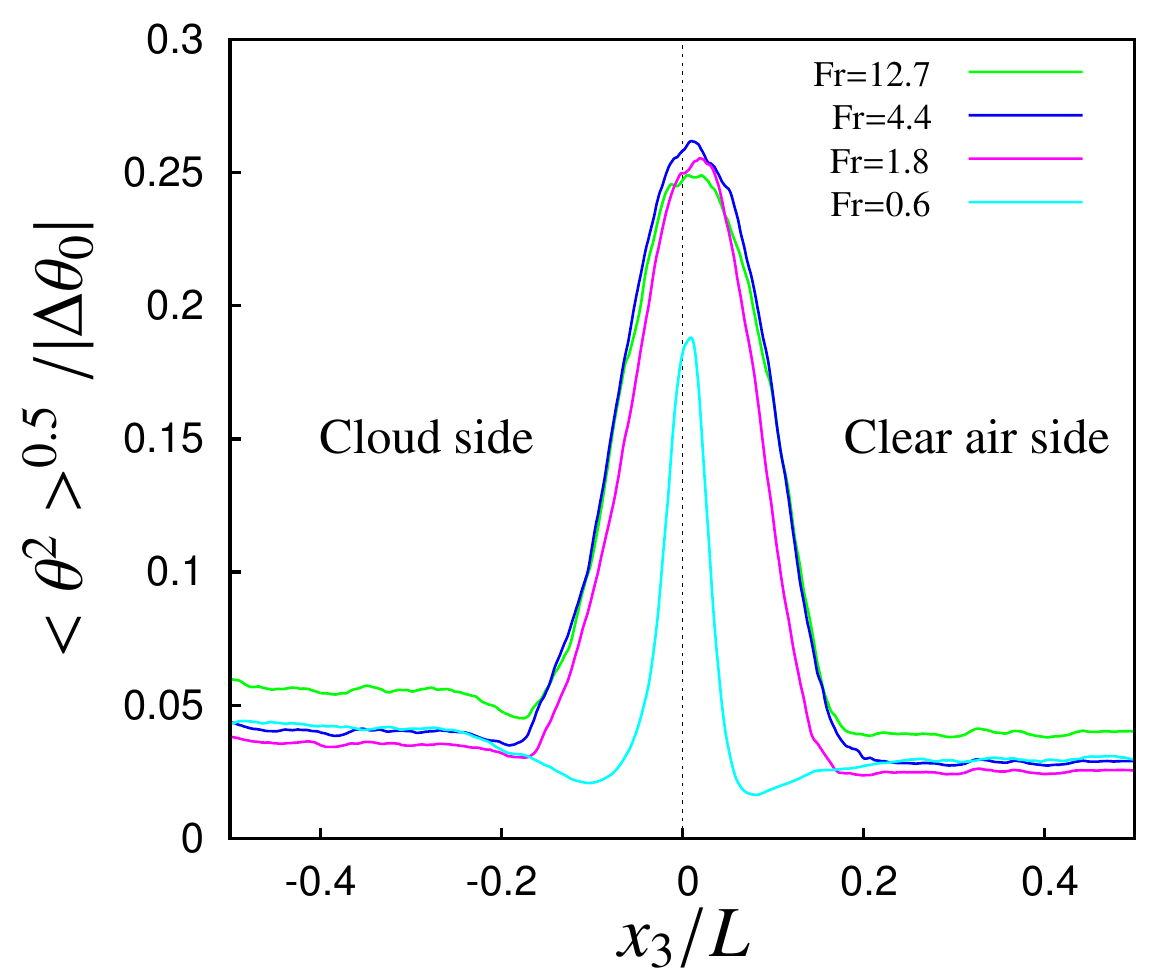}}
\begin{minipage}{.49\textwidth}
	\subfigure[]{\includegraphics[width=\textwidth]{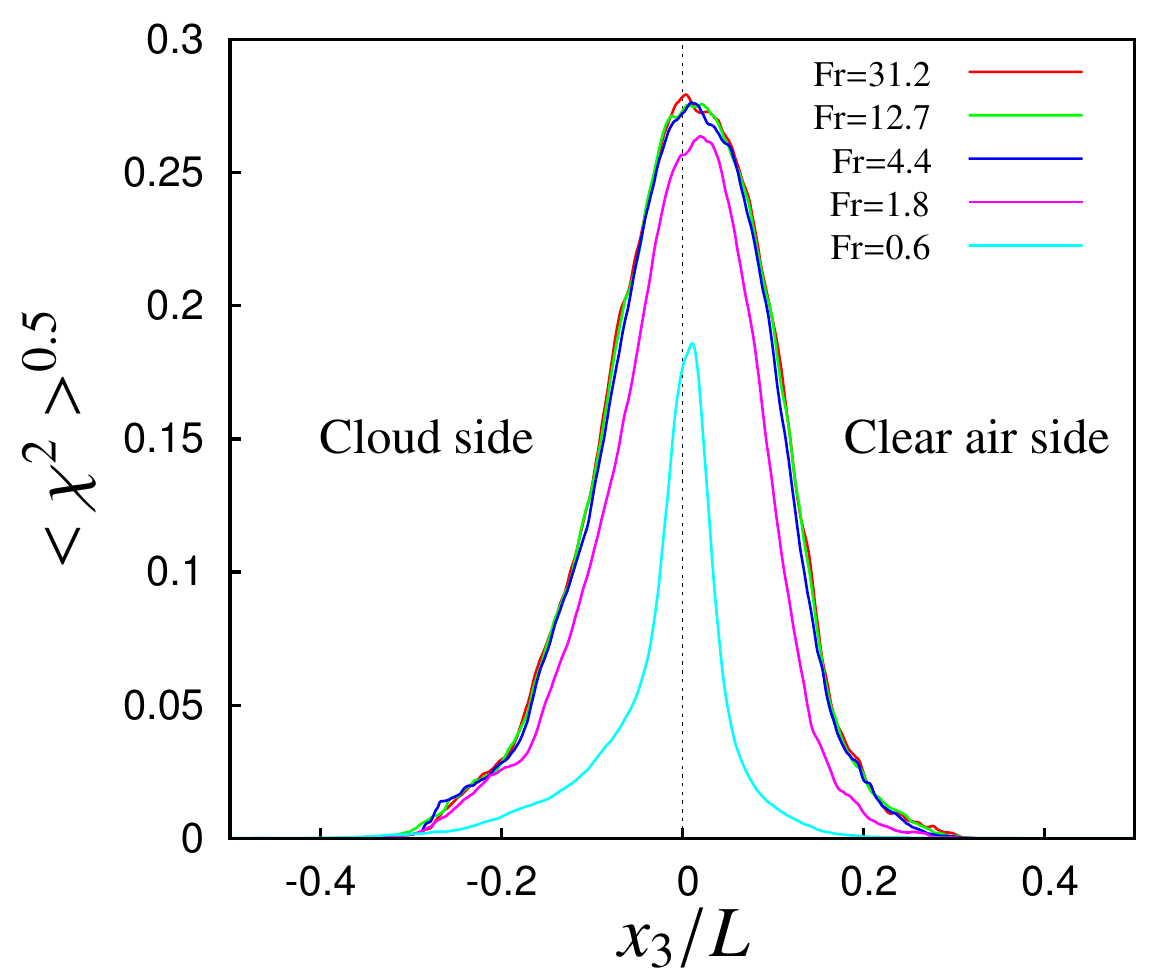}}
\end{minipage}
\hfill
\begin{minipage}{.49\textwidth}
	\caption{Spatial distribution along the vertical coordinate $x_3$ of the kinetic energy (a), temperature variance (b) and water vapor concentration variance (c) at $t/\tau\approx6$ for different levels of stratification.}
\end{minipage}
	\label{fig:var}
\end{figure}

 \begin{figure}[p]
 	\centering
 	\subfigure[]{\includegraphics[width=.49\textwidth]{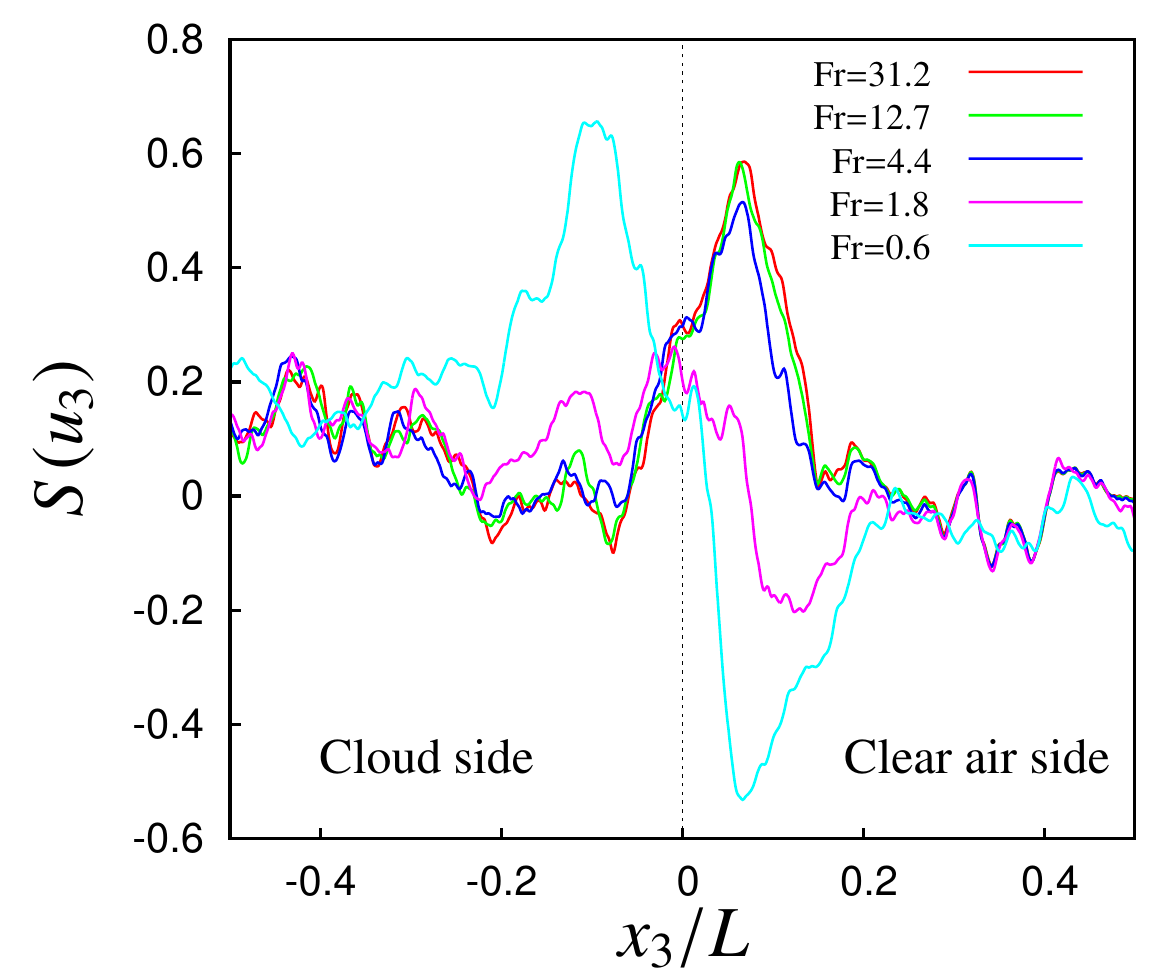}}
 	\hfill
 	\subfigure[]{\includegraphics[width=.49\textwidth]{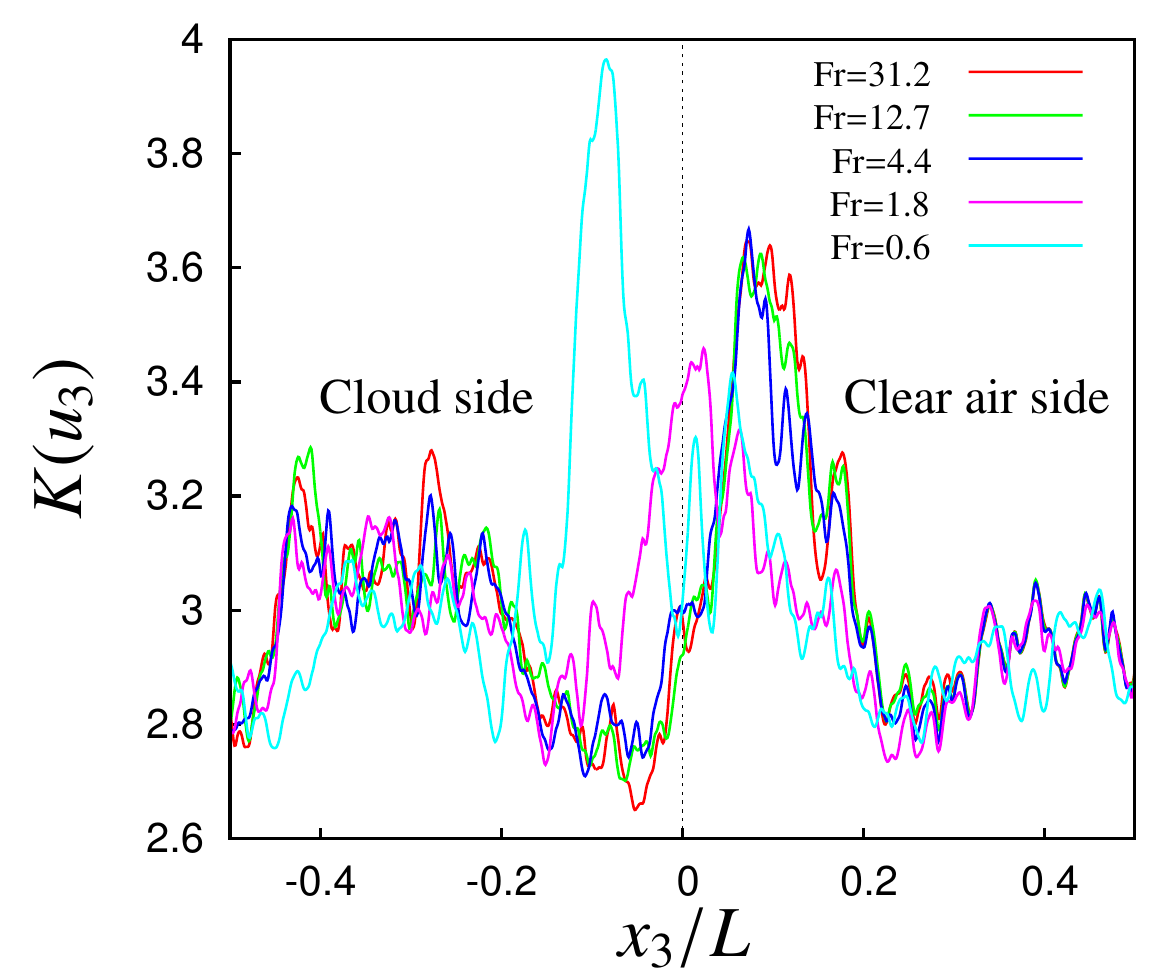}}
 	\subfigure[]{\includegraphics[width=.49\textwidth]{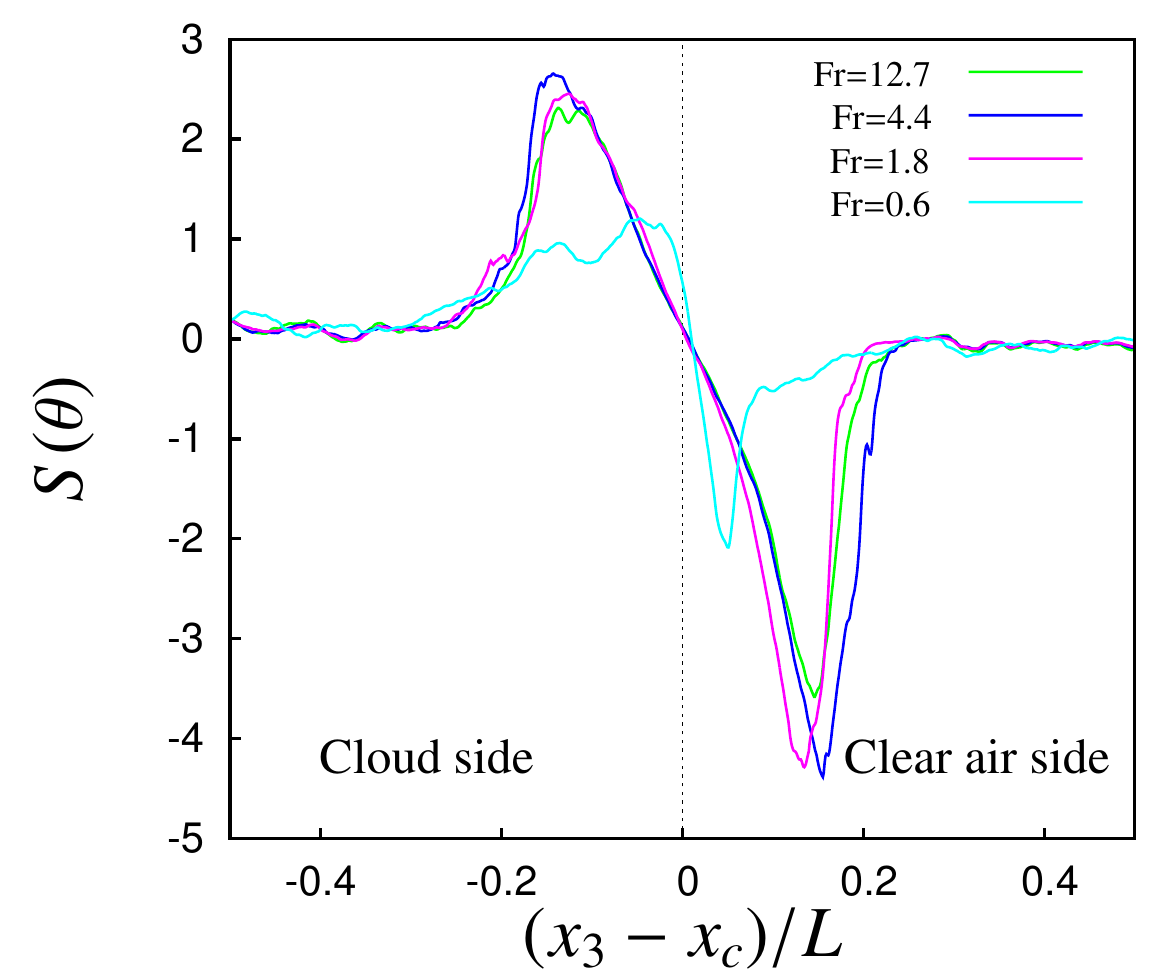}}
 	\hfill
 	\subfigure[]{\includegraphics[width=.49\textwidth]{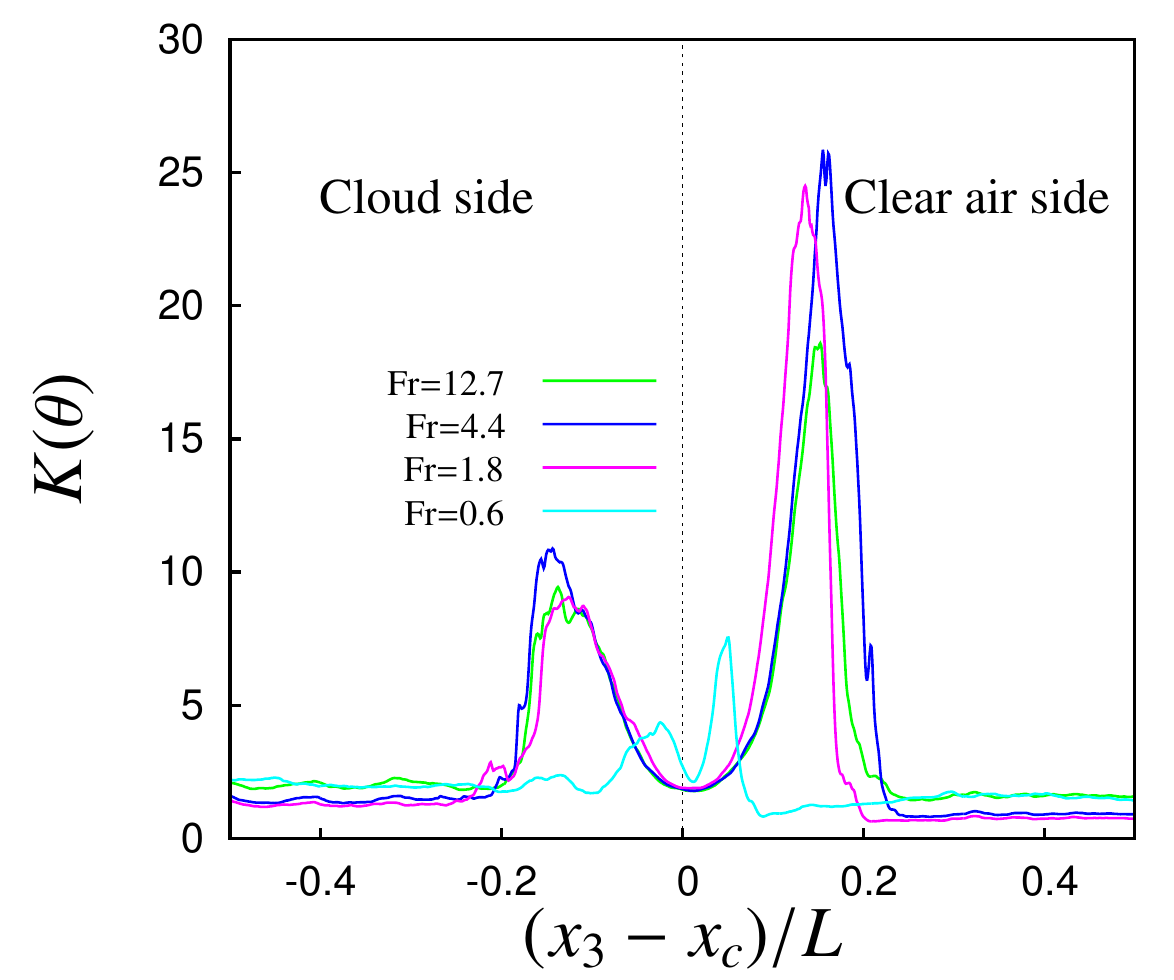}}
 \subfigure[]{\includegraphics[width=.49\textwidth]{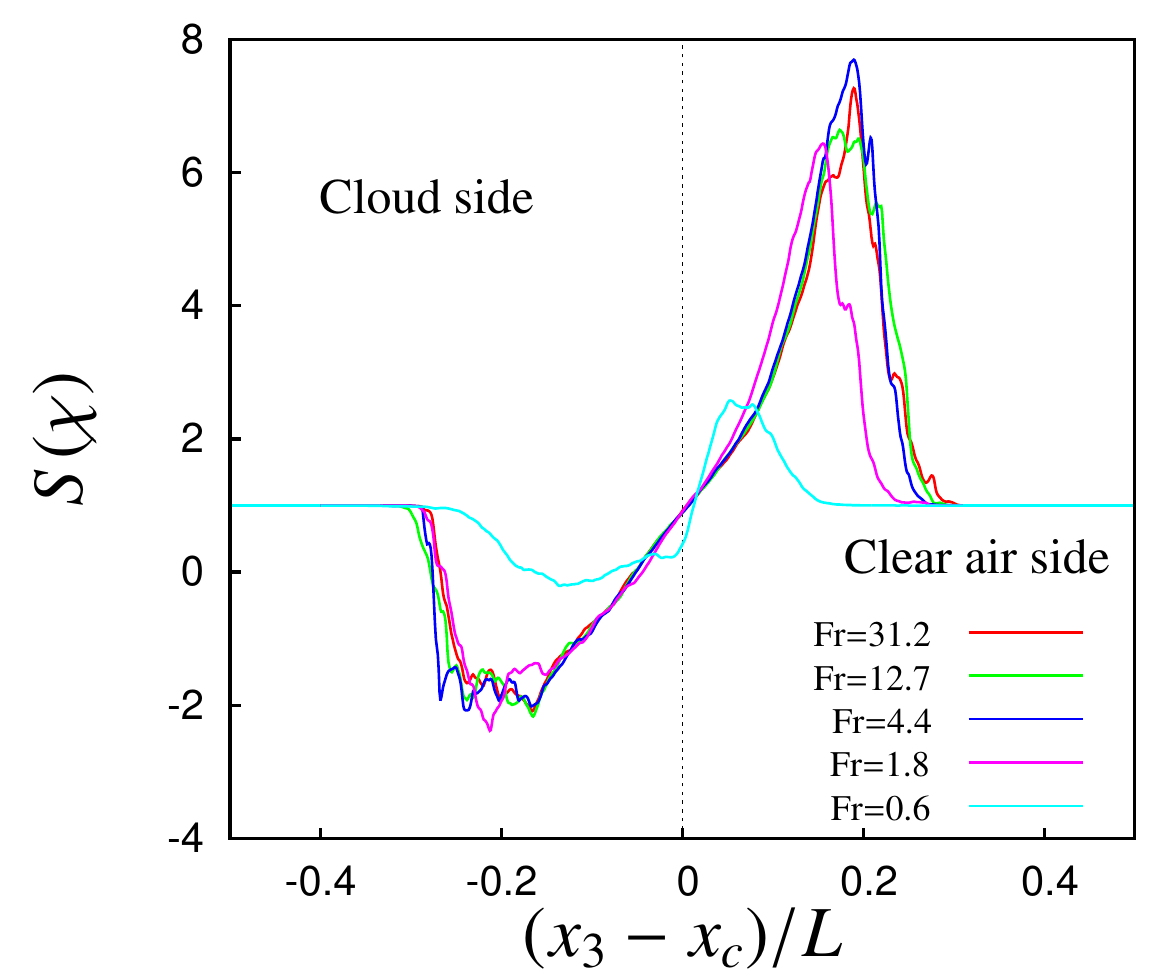}}
 	\hfill
 \subfigure[]{\includegraphics[width=.49\textwidth]{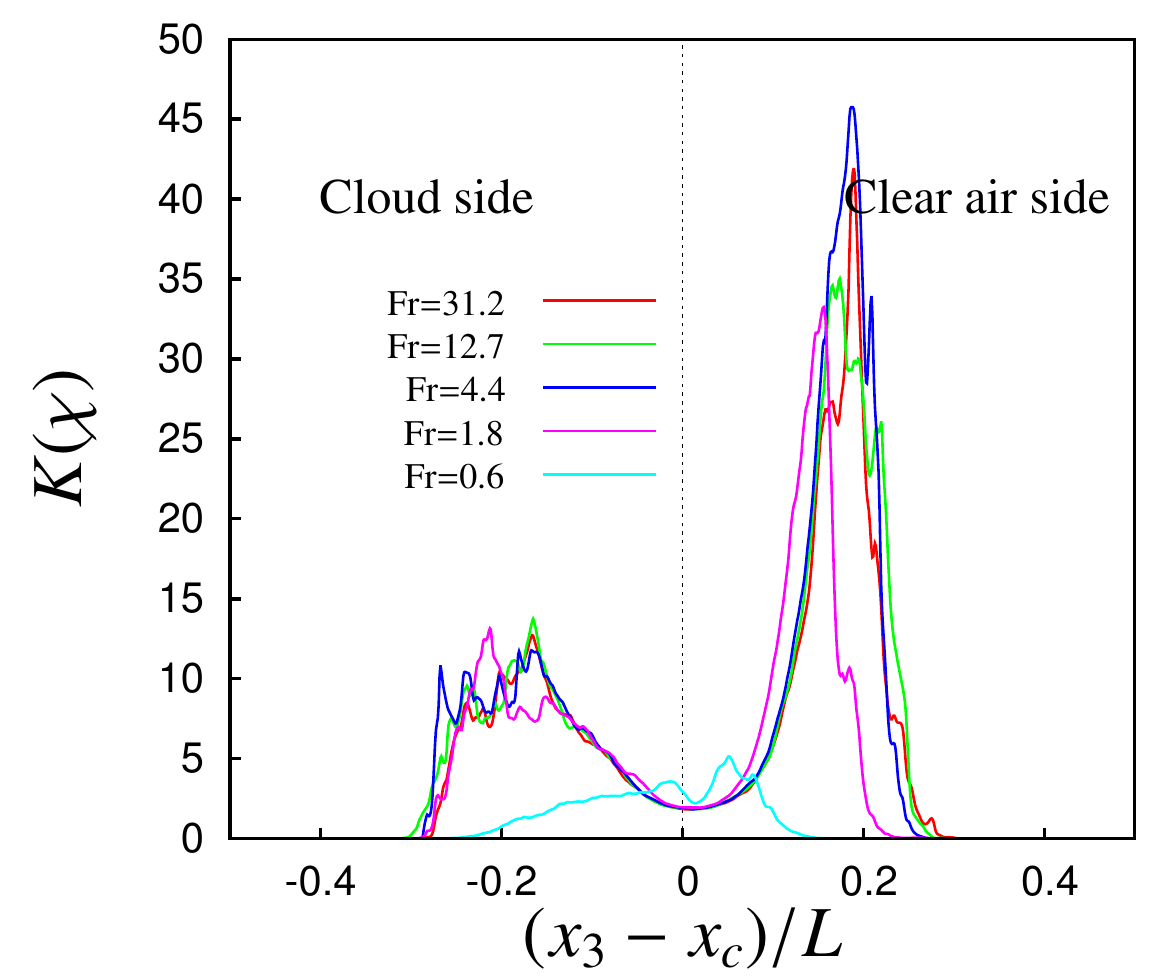}}
 	\caption{Spatial distribution in the vertical direction $x_3$ of the skewness and the kurtosis of vertical component of the velocity (a - b), temperature (c - d) and water vapour concentration (e - f) at $t/\tau=6$ for different levels of stratification.}
 	\label{fig:kur}
 \end{figure}
According to the ratio between buoyancy force and kinematic forces (that are advection and diffusion), the evolution of the system can be split in  two main stages.
As long as the ratio remains small, there are no significant differences with respect to a non-stratified case.
On {the} contrary, as the stratification perturbation level become higher, buoyancy effects are no more negligible: differences are present from both a quantitative and qualitative point of view.
These considerations can be observed {through the statistical analysis of the simulated fields.  The statistics are computed by averaging the variables in the planes normal to the mixing direction (with a sample of $1024\times1024$ data-points), focusing on the variation along the vertical (non-homogeneous) direction.
The effects of the different stratification levels are clearly visible on the second order moment of velocity, temperature, and vapor as shown in figure \mbox{\ref{fig:var}} (a - b - c).  
When the stratification level is mild ($Fr>4$) there are no relevant differences with respect the neutrally buoyant flow, while significant differences appears for intense stratification ($Fr<2$)}. In particular, in correspondence of the local temperature perturbation, the formation of a layer with a pit of kinetic energy can be observed. 
The presence of such a layer deeply changes the physics of the system, because in this situation two interfaces are produced. The first interface, (which is present also in the absence of stratification), now separates the high turbulent energy region from the pit, while the second one (not present without stratification) separates the low turbulent energy region from the center of the mixing layer. Therefore, a strong stratification induces a physical separation between the two external regions, greatly decreasing the interaction between them. 
Both interfaces present an intermittent behavior, as shown in figure \ref{fig:kur} (a - b) by skewness and kurtosis distribution {(respectively third and fourth moments normalized with the local variance)}. In fact, two peaks of skewness and kurtosis can be observed in the highly stratified case: one is placed inside the cloud, and the other is placed close to the position of the intermittency peak in case of absence of stratification, see data at $Fr=0.6$ in figure \ref{fig:kur} (a).
Observing the magnitude of the kurtosis maximum in figure \ref{fig:kur} (b), it can be noted that the peak inside the cloud reaches values as high as 4, that are about the 10\% larger than when the stratification is milder.

The behavior of the temperature and the water vapor fields appear qualitatively analogue as far as the shape of the statistical distributions are concerned, but are quantitatively different. See, for instance, the temperature and the vapor concentration peaks in the skewness and kurtosis distribution, shown in figure \ref{fig:kur} (c - d - e - f). 



It can be also observed that higher levels of stratification produce a relevant reduction of intermittency in the flow, with a {drop} of about 70\%\ in the peaks of skewness and kurtosis. The interaction between the two regions aside the interface is greatly reduced, so the fluctuation at the sides of the mixing layer are damped, preventing the formation of the intermittent layer typical of the passive scalar transport \cite{mw86,jturb}. 
{This strong reduction in skewness and kurtosis is coupled with a slight increase in the higher order moments of vertical velocity -- for sufficiently strong stratification -- which is in fair agreement with the trend observed in \mbox{\cite[p. 158]{qiu08}}}.
Moreover, observing both figures \ref{fig:var} and \ref{fig:kur}, it is clear that the thickness of the mixing layer is reduced in case of intense stratification (see next section for more details).

\subsection{The onset of a kinetic energy pit}

\begin{figure}
	\centering
	\subfigure[]{\includegraphics[width=.49\textwidth]{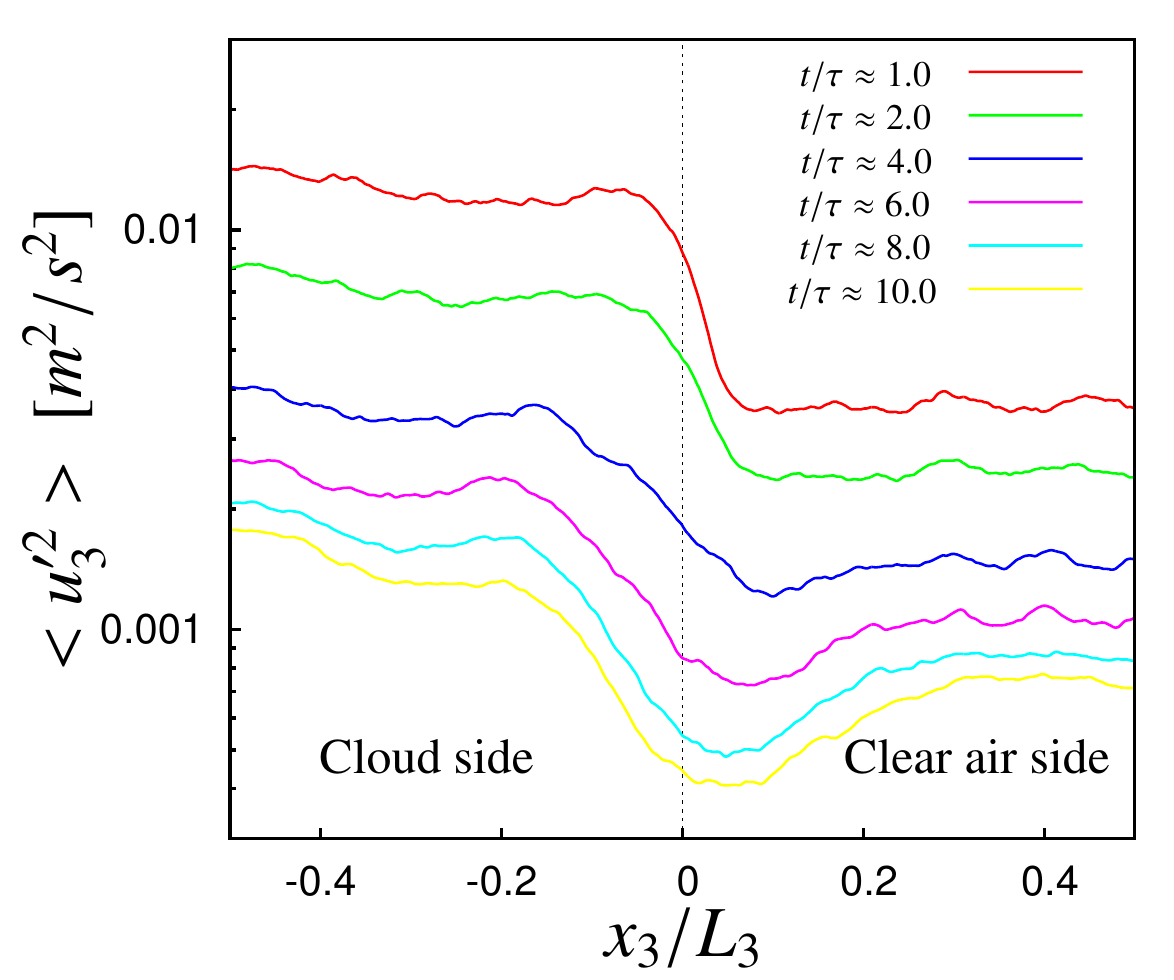}}
	\hfill
	\subfigure[]{\includegraphics[width=.49\textwidth]{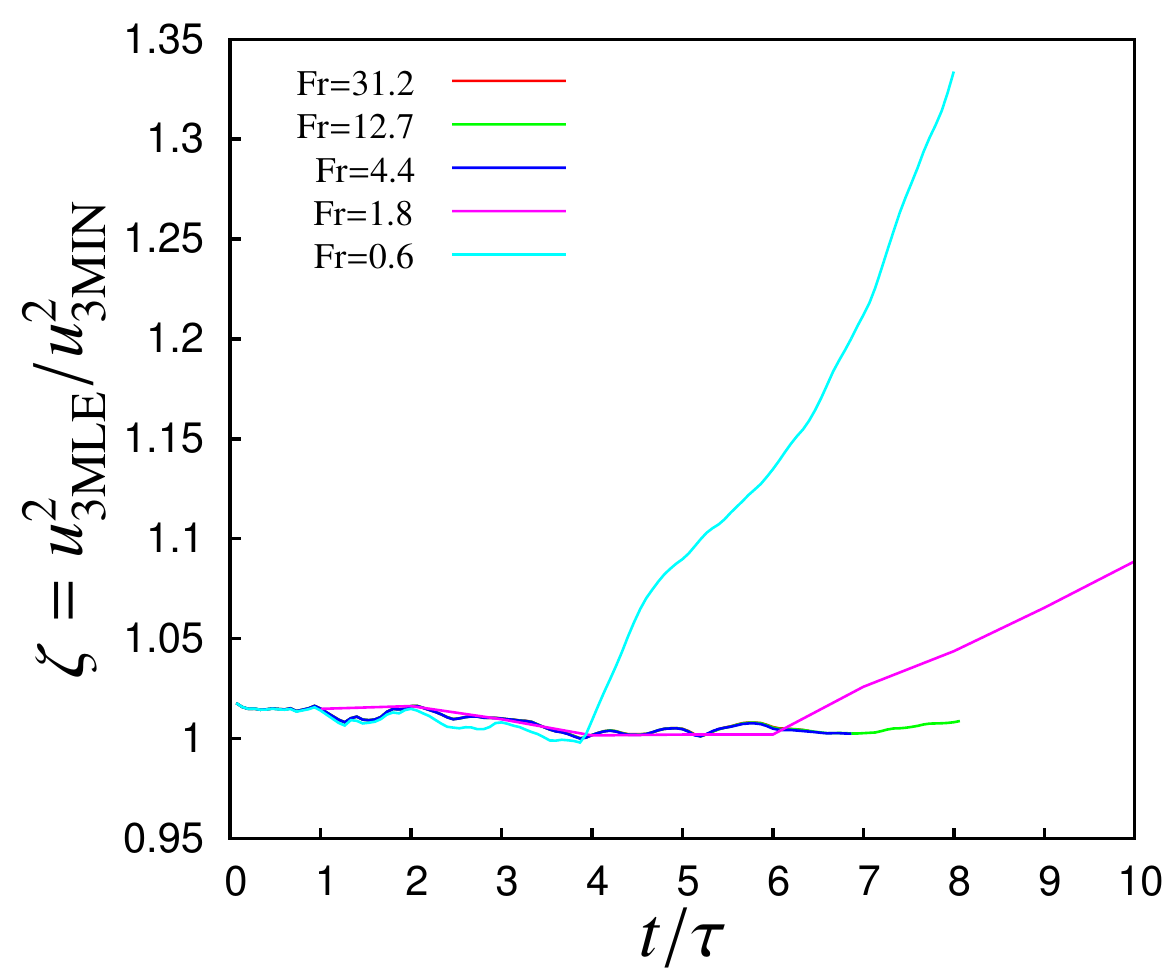}}
	\caption{
	(a) Temporal evolution of the spatial distribution along the vertical direction $x_3$ of the turbulent kinetic energy when Fr=$1.8$. A pit of energy appears after about 6 eddy turnover times in correspondence of the original interface.
	(b) Temporal evolution of the ratio $\zeta$ between the mean vertical velocity variance in the lower energy (clear air side) region $u_{3,MLE}$ and the minimum value of the vertical velocity $u_{3,MIN}$. When this ratio departs from 1, a pit of kinetic velocity appears as shown in (a).}
	\label{fig:pit}
\end{figure}
\begin{figure}
\centering
\subfigure[]{\includegraphics[width=.49\textwidth]{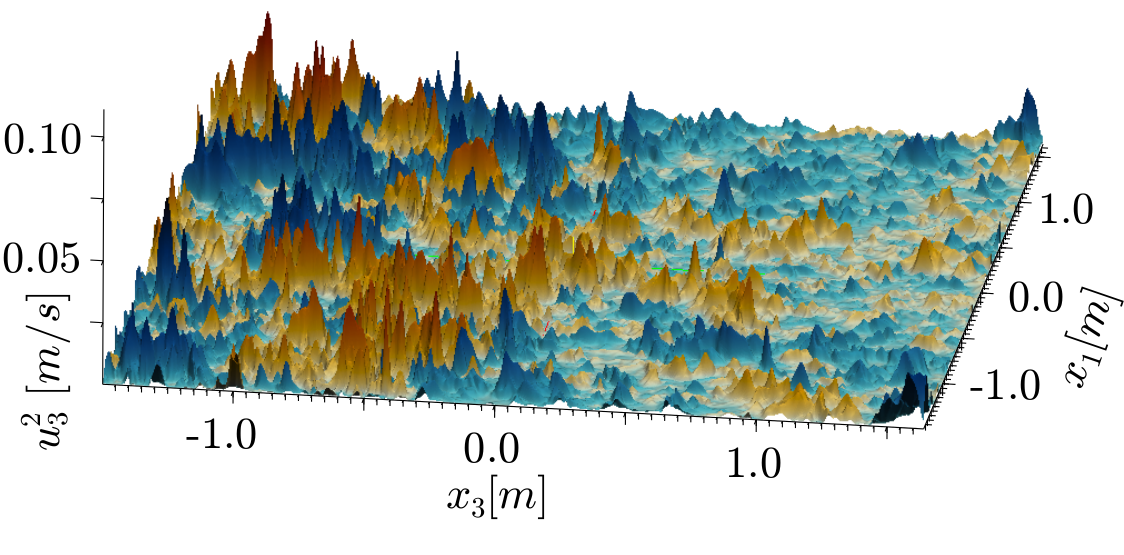}}
\subfigure[]{\includegraphics[width=.49\textwidth]{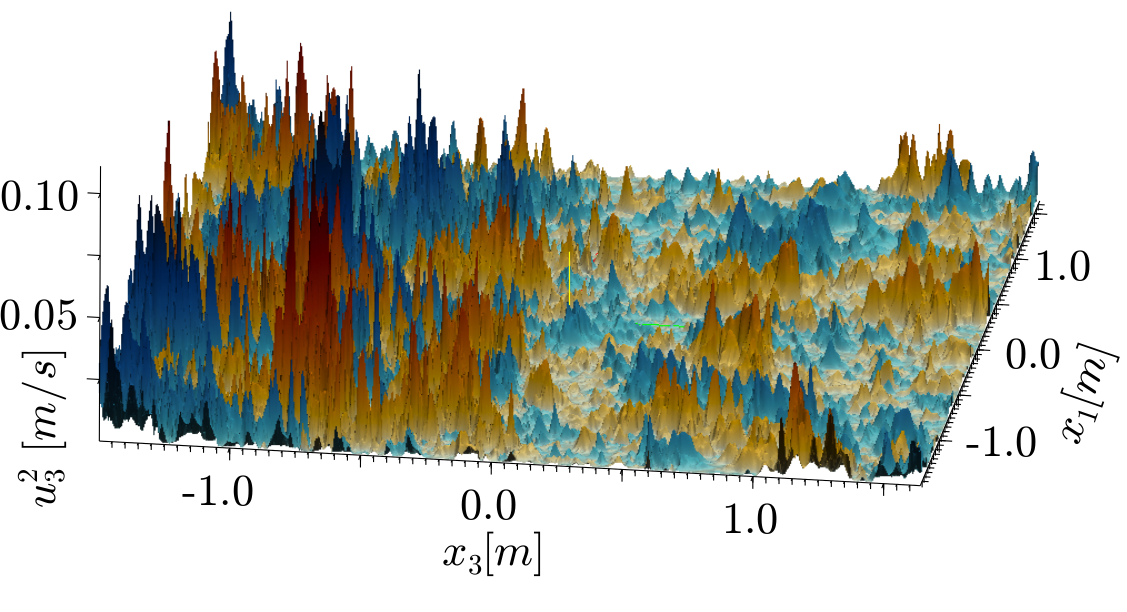}}
\begin{minipage}{.49\textwidth}
\subfigure[]{\includegraphics[width=\textwidth]{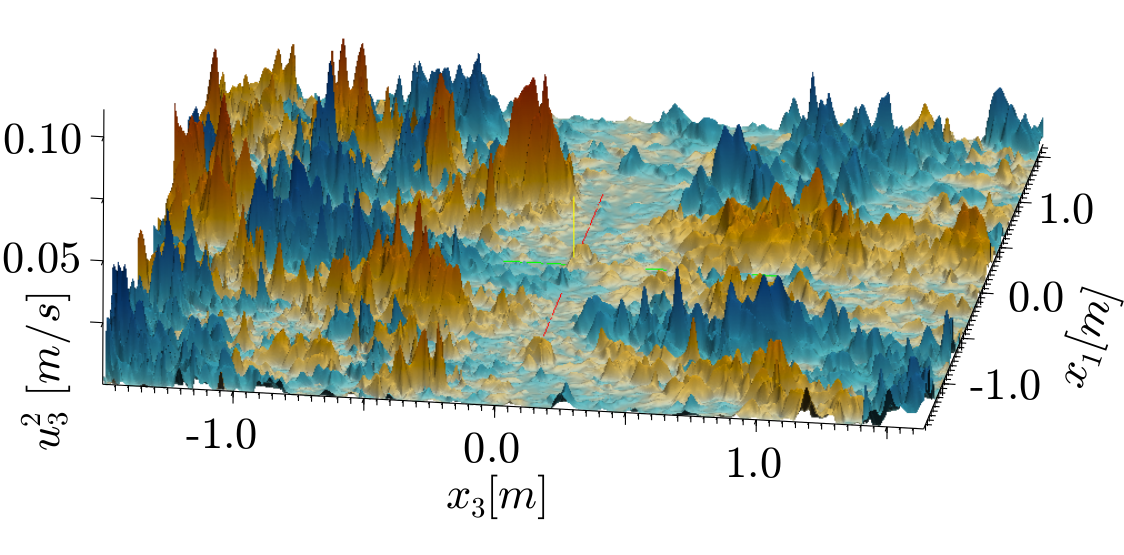}}
\end{minipage}
\begin{minipage}{.49\textwidth}
\caption{\label{fig:elev2}Visualization of the vertical component of the velocity in a vertical plane. The elevation is proportional to the square of vertical velocity fluctuations and the colors respect the velocity directions (blue for downward , red for upward) in various case: $Fr=12$ after 8 eddy turnover times (a) and for Fr=$0.62$ after 4 and 8 eddy turnover times (panels b, c respectively). In the last two panels it can be observed the formation of the pit of kinetic energy.}
\end{minipage}	
\end{figure}

As shown in figure \ref{fig:pit} (a), in case of high stratification level, in the center of the domain {-- where the initial temperature step is placed} -- the onset of a layer with a kinetic energy lower than both the external regions can be seen. {This layer can be considered} as a pit of kinetic energy.
Varying the stratification intensity, the genesis and the evolution of such pit can be measured by considering the ratio $\zeta$ between the variance of vertical velocity in the low energy region {$u^2_{3MLE}$} (mean value) and in the center of the pit {$u^2_{3MIN}$} (where the variance reaches its minimum); {the temporal evolution of $\zeta$} is shown in figure \ref{fig:pit} (b).
{Qualitatively similar results have been observed in the large-eddy simulations of stratocumulos-topped planetary boundary layer carried out by several physics of the atmosphere research group, as described in \mbox{\cite[fig.7(c) and 8(a) at pag. 11]{moe1996}}:
in particular, in the case of sufficiently strong stratification, the trend of our  vertical velocity variance,
in figure \mbox{\ref{fig:entr}}(a), is analogue to those observed in the LES carried out by the NCAR group
 (Deardoff TKE model \mbox{\cite{moeng86}}) and the WVU (ARAP TKE model \mbox{\cite{sykes}}). In this simulations they consider the planetary boundary layer with $Re_\lambda\approx5500$ and $Fr\approx0.4$.}


A visualization of such phenomenology is represented in figure \ref{fig:elev2}, where the vertical velocity fluctuations in a vertical slice of the domain are represented using an elevation plot (where such elevation is proportional to the square of $u_3$). In presence of a mild stratification, $Fr=12$, even after 8 time scale, there is a smooth mixing layer between the high (left) and the low (rigth) energy regions. 
The differences in the case of strong stratification, with $Fr=0.62$, after 4 time-scales (b) and 8 time scales (c), are clearly visible: a separation layer is present in the center of the domain, that becomes even more evident as the time pass by.

\begin{figure}
	\centering
	\subfigure[]{\includegraphics[width=.49\textwidth]{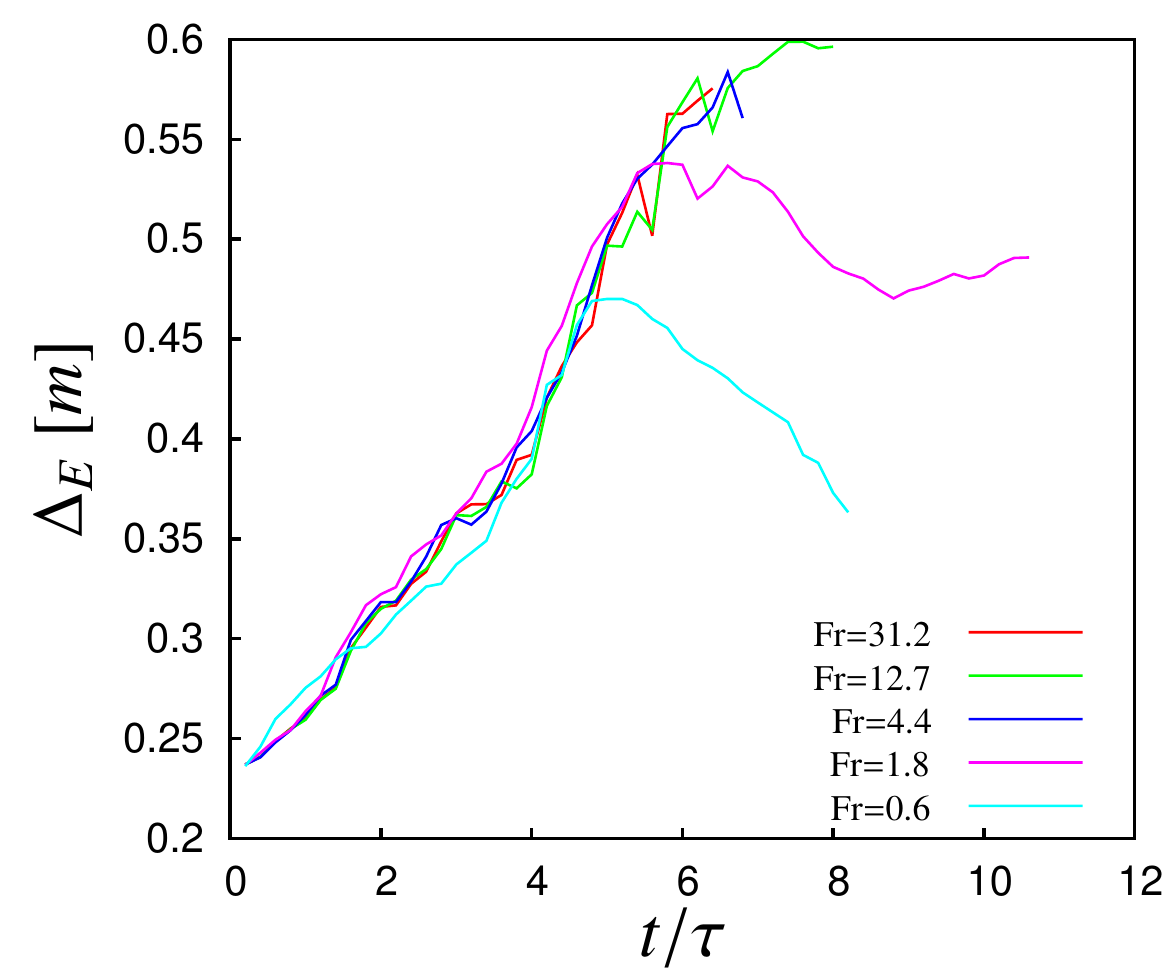}}
	\hfill
	\subfigure[]{\includegraphics[width=.49\textwidth]{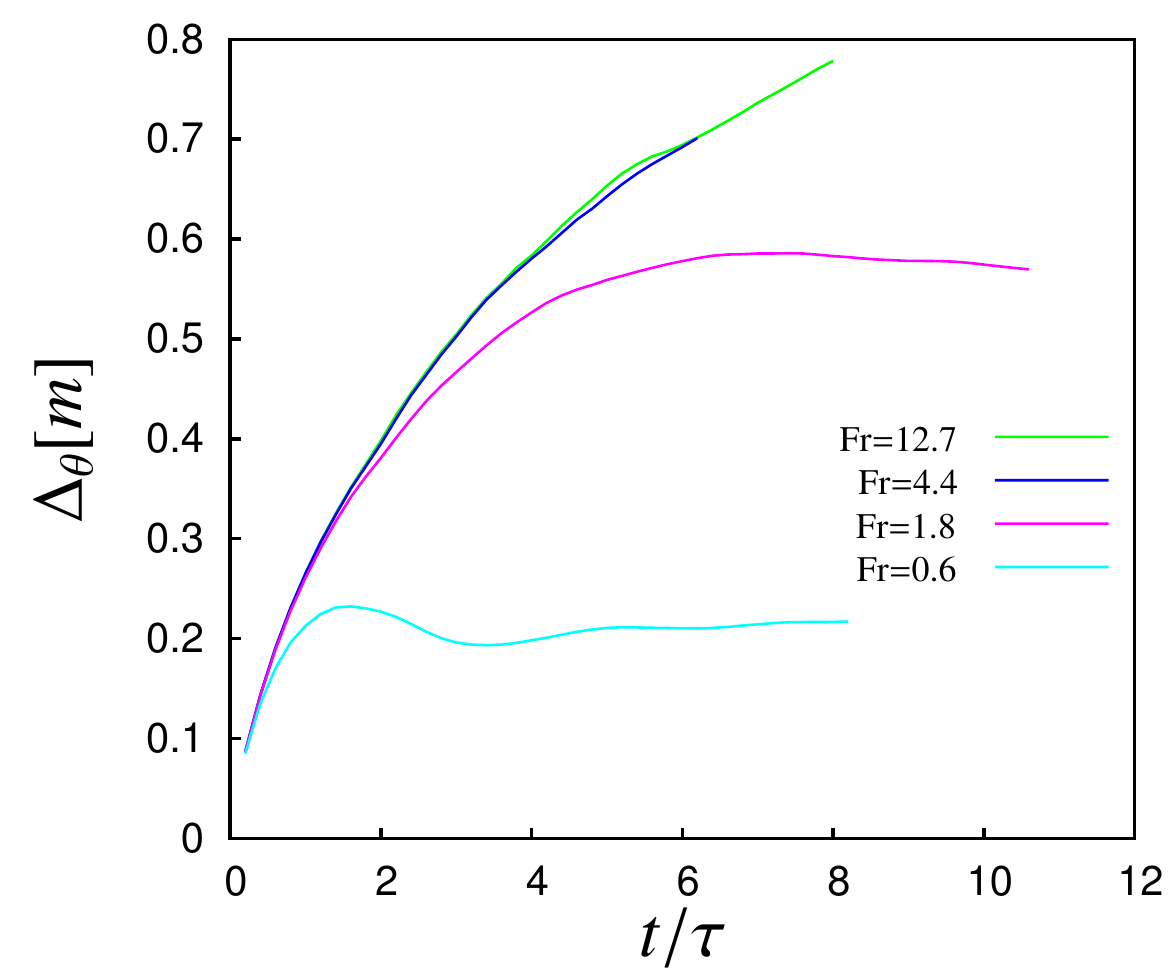}}
	\caption{Thickness of the mixing layer of velocity (a) and temperature (b). For both fields, the mixing layer thickness is the same as defined in \cite{jturb}.}
 	\label{fig:thick}
\end{figure}

{As said, the} presence of the pit generates a physical separation between the two external regions, by damping the turbulent mixing, and thus reducing the exchange of information. As a consequence, there is a saturation of the thickening of the mixing layer $\Delta_E$; {such interruption of the growth is represented in   figure \mbox{\ref{fig:thick}} (a)}

Looking to the temperature mixing layer thickness $\Delta_\theta$, shown in figure \ref{fig:thick} (b), it can be seen that,  for strong stratification, the thickening stops approximately after the same amount of times scale required by $\Delta_E$. In that case, contrary to what observed for the kinetic energy, the thickening does not stop suddenly, but rather with a transient that lasts a couple of time scales.

\subsection{Entrainment}

The entrainment of clear air inside the cloud is an important aspect of the top cloud interface as it concurs in the evaporation/condensation of droplets inside a cloud \cite{wo12}.

\begin{figure}
\vskip -4eX
	\centering
	\subfigure[]{\includegraphics[width=.49\textwidth]{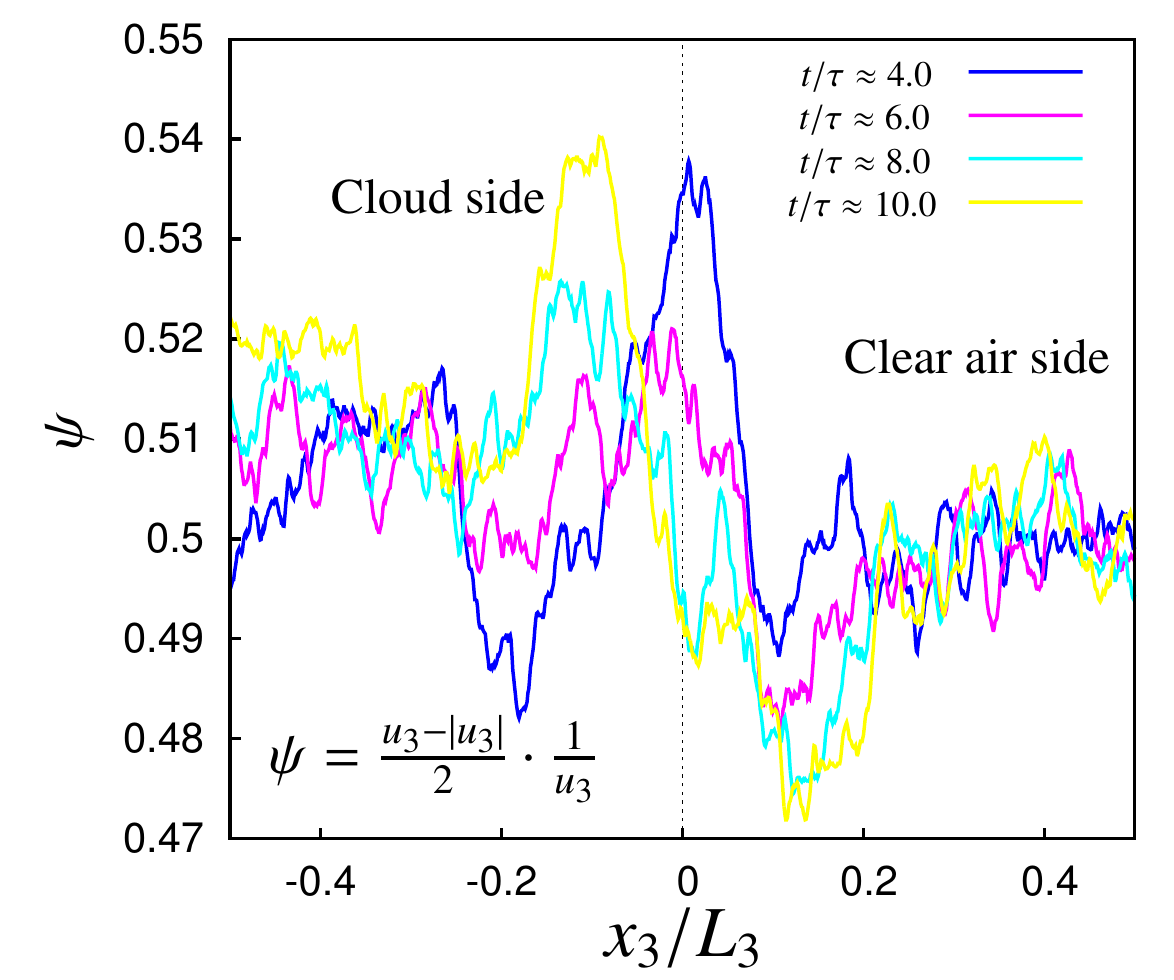}}
\hfill	
  \subfigure[]{\includegraphics[width=.49\textwidth]{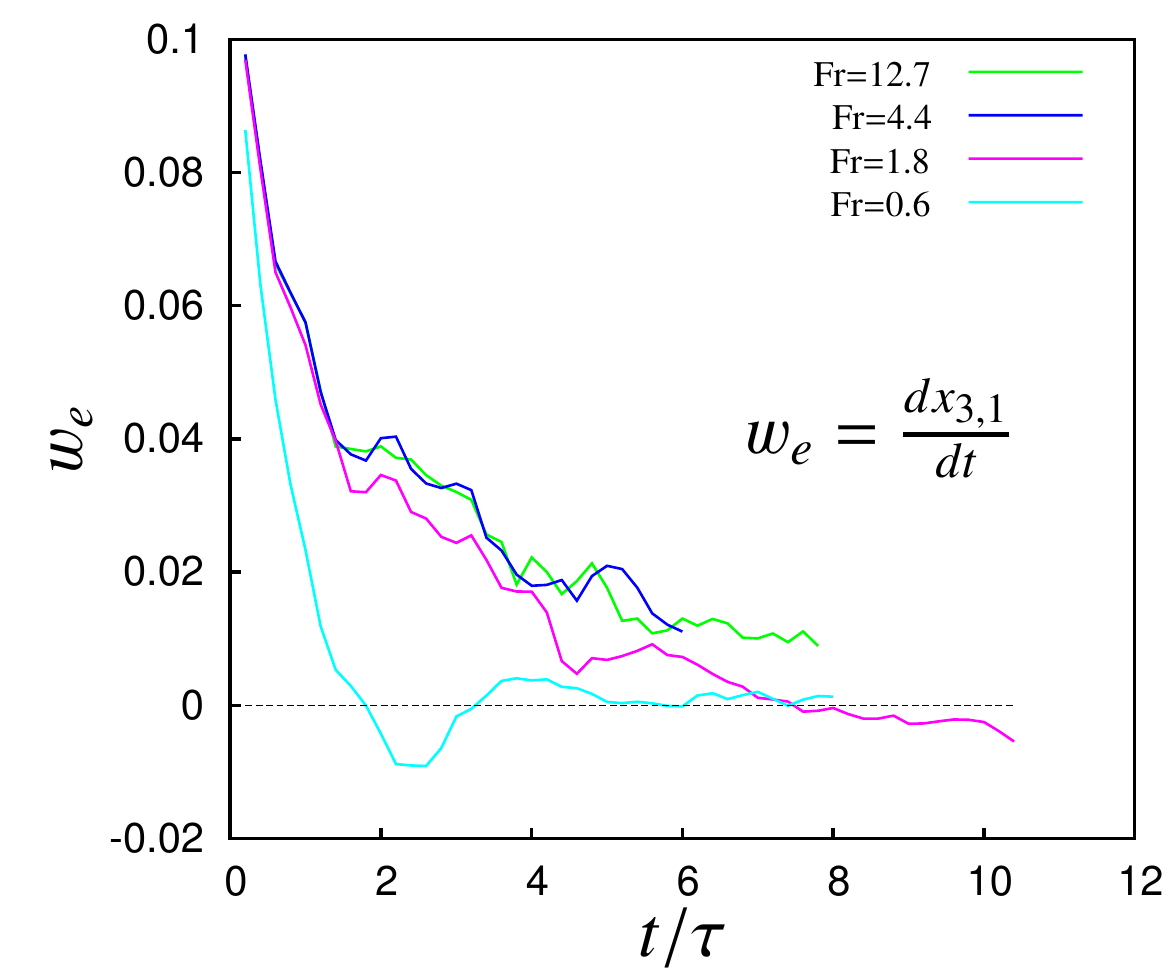}}
	\begin{minipage}{.49\textwidth}
	\subfigure[]{\includegraphics[width=\textwidth]{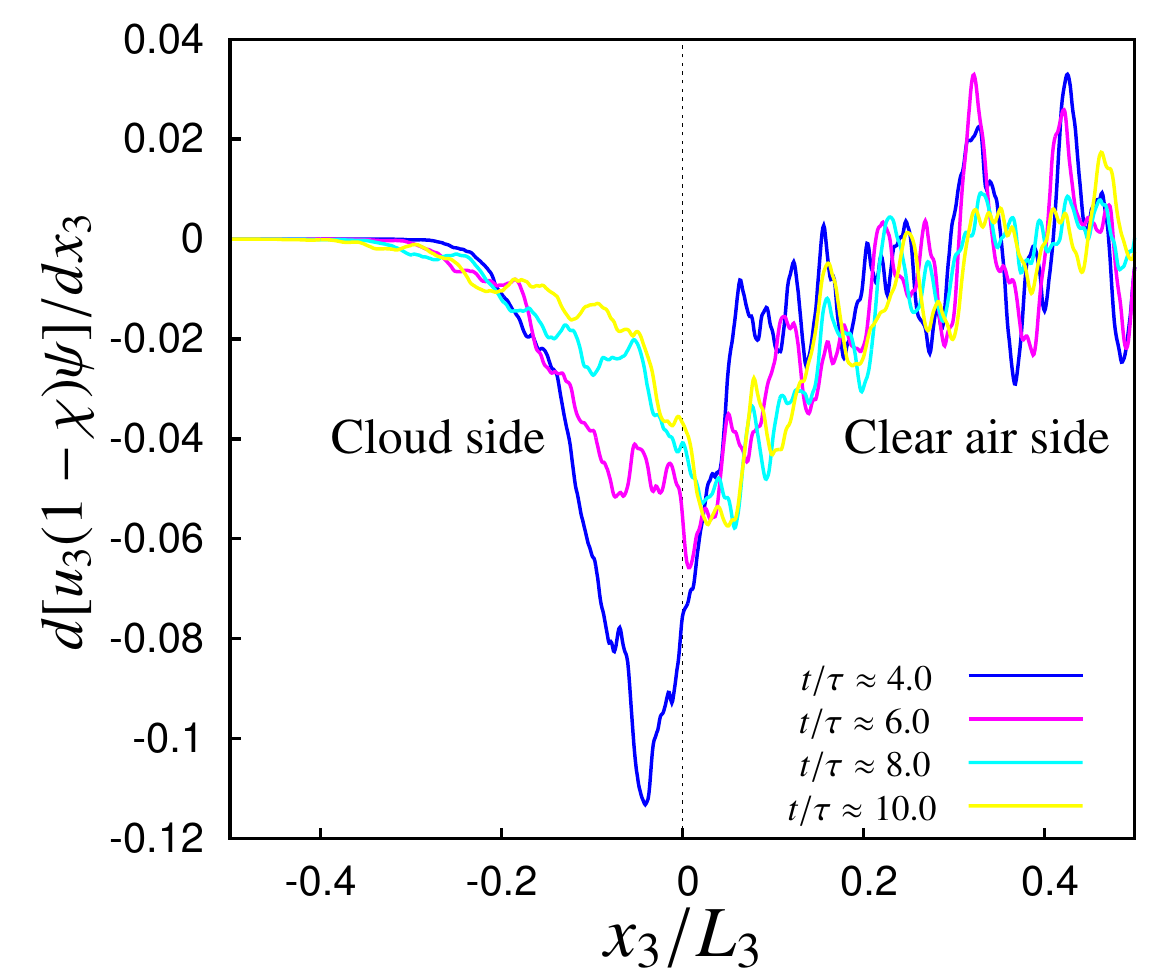}}
	\end{minipage}
	\hfill
	\begin{minipage}{.49\textwidth}
	\caption{\label{fig:entr}Evolution of the entrainment  across the top cloud. Fraction $\psi$ of downwards velocity (a) and vertical variation of the mean flux of clear air into the cloud (c) when with Fr$=1.8$
	\newline
	Evolution of entrainment velocity $w_e$ normalized with the high kinetic energy $E_1$ root mean square (b), see figure \ref{fig:schema}.}
	\end{minipage}
\end{figure}

In any plane parallel to the interface, in absence of a mean velocity, only downward velocity fluctuations can transport clear air into the cloud. Their presence can be represented by a marker function $\psi$ that is equal to 1 where $u_3$ is negative, and 0 otherwise. Its average in each horizontal plane, shown in figure \ref{fig:entr} (a), shows a small deviation from the mean value of 0.5 which would be observed in an homogeneous and isotropic flow. This implies that upward and downward fluctuations are almost equally distributed; the maximum departure from an homogeneous distribution is about 4\%, and the spatial distribution of $\psi$ looks like the one which has been observed in the third order moment of the velocity, see figure \ref{fig:kur} (a).

The entrainment of clear air is responsible of the growth of the cloud. In fact, the velocity $w_e=dx_{3,i}/dt$ of the cloud top interface ($\left\langle x_{3,i}\right\rangle$ is the mean vertical position of the cloud top, defined as the position where the mean vapor concentration is 25\%) has often been used as a parameter to measure the entrainment rate \cite{mel2010,moe2000}.
In figure \ref{fig:entr} (b), the temporal decay of $w_e$ for different levels of stratification {is represented}. In presence of weak stratification, that is $Fr$ larger than 4, its value gradually decreases with an almost exponential law, due to the decay of the kinetic energy. On the contrary, when the stratification is stronger, the damping of $w_e$ is much faster, and the entrainment vanishes after few times scale, when the presence of the pit of kinetic energy substantially reduces the flux of clear air inside the cloud.

Figure \ref{fig:entr} (c) shows the vertical derivative of the downward flux of clear air when $Fr=1.8$
As the flow evolves, the downward flux reduces and its derivative, which represents the net variation of $1-\chi$ at a given instant, rapidly tends to zero inside the cloud. This implies that the entrainment of clear air is confined to a thin interfacial layer.

\section{Conclusion}
In this work we have carried out numerical simulations on the transport of energy and scalars in a turbulent shearless mixing layer associated to temporal perturbation of the temperature lapse rate across the clear air - cloud interface. The perturbation locally introduces a stable stratification. 
This idealized configuration models some of the phenomena which are present in the kinetic dynamics of the cloud and clear air interaction, namely those linked to turbulent mixing and entrainment.

We have shown that this flow configuration develops an horizontal layered structure characterized by a sublayer -- with a kinetic energy lower than both the external regions -- which acts as a barrier for the transport between the cloud and the external ambient. In our transient simulations, this flow structure appears when the buoyancy terms becomes of the same order of magnitude of the inertial ones, therefore the time needed for this transition becomes shorter when the stratification is more intense.
Once buoyancy dominates and the new flow regime is reached, we observed two highly intermittent regions with opposite kinetic energy gradients.
As a direct consequence, the entrainment is damped.

{Results obtained so far seemingly support the large eddy simulations of stratocumulus - topped planetary boundary layers}.

%
\section*{Bibliography}
 \bibliography{strat_bib}

\end{document}